\documentclass[a4paper, 10pt, conference]{IEEEtran}
\IEEEoverridecommandlockouts
\usepackage{todonotes}
\usepackage[hyphens]{url}

\usepackage{hyperref}

\usepackage{mathtools}
\usepackage{amsmath}
\usepackage{amsthm}
\usepackage{amssymb}
\usepackage{float}
\usepackage{graphicx}
\usepackage{subcaption}
\usepackage{multicol}
\usepackage{paralist}
\usepackage{bm}
\usepackage{amsfonts}
\usepackage{url}
\usepackage{multirow}
\usepackage{cases}
\usepackage{booktabs}
\usepackage{color}
\usepackage{hyperref}
\usepackage{balance}
\hypersetup{
  colorlinks   = true, 
  urlcolor     = black, 
  linkcolor    = blue, 
  citecolor   = black 
}

\theoremstyle{definition}

\usepackage{amsmath}

\begin{document}
\title{Adult content in Social Live Streaming Services: Characterizing deviant users and relationships}

\author{
\IEEEauthorblockN{Nikolaos Lykousas\IEEEauthorrefmark{1},  Vicen\c{c} G\'omez\IEEEauthorrefmark{2}, Constantinos Patsakis\IEEEauthorrefmark{1}}
\IEEEauthorblockA{\IEEEauthorrefmark{1}University of Piraeus, Greece \url{nikos.lykousas@gmail.com},\url{kpatsak@unipi.gr}}
\IEEEauthorblockA{\IEEEauthorrefmark{2}Universitat Pompeu Fabra. Barcelona, Spain \url{vicen.gomez@upf.edu}}
}


\maketitle
\begin{abstract}
Social Live Stream Services (SLSS) exploit a new level of social interaction.
One of the main challenges in these services is how to detect and prevent deviant behaviors that violate community guidelines.
 In this work, we focus on adult content production and consumption in two widely used SLSS, namely Live.me and Loops Live, which have millions of users producing massive amounts of video content on a daily basis. We use a pre-trained deep learning model to identify broadcasters of adult content.
Our results indicate that moderation systems in place are highly ineffective in suspending the accounts of such users.
We create two large datasets by crawling the social graphs of these platforms, which we analyze to identify characterizing traits of adult content producers and consumers, and discover interesting patterns of relationships among them, evident in both networks.
\end{abstract}

\section{Introduction}
\noindent 
The wide adoption of mobile technologies have completely redesigned the way we consume and produce information as well as the way we interact with people.
This shift and cultural change has 
lead to the emergence of many new Social 
Media platforms that focus on features and topics that the traditional ones like Facebook and Twitter are lacking, with typical examples including Snapchat, Periscope, and musical.ly. Many of these platforms 
 operate solely on mobile devices.

Social Live Streaming Services (SLSS) are examples of this new type of platforms in which 
users can actually live stream parts of their daily lives. These services provide a new level of interaction and hook their subscribers as the users become part of the daily life of others. Practically, users decide when to open up their cameras and share snapshots of what they do, what they think or live at the moment with others and interact with them via chat messages.

In this work, we analyze a gray area of these services: adult content production and consumption. Clearly, most SLSS have a clear policy against adult content and facilitate some mechanisms to detect and ban the misbehaving users, either in the form of filters from the service provider or by peer reporting. 

To this end we consider two SLSS, namely, Live.me (\url{www.liveme.com}) and Loops Live (\url{www.loopslive.com}), from now on LM and LL, respectively. Both operate as video chat apps solely in mobile phones and have millions of users that produce massive amounts of video content on a daily basis. To quantify the latter, Cheetah Mobile's CEO; the company which owns Live.me, reported that more than 200,000 hours of live video are broadcast daily on Live.me\footnote{\url{https://seekingalpha.com/article/4075406-cheetah-mobiles-cmcm-ceo-fu-sheng-q1-2017-results-earnings-call-transcript}}. Both platforms are very successful, especially in young users\footnote{\url{https://seekingalpha.com/article/4025223-cheetah-mobiles-cmcm-ceo-fu-sheng-q3-2016-results-earnings-call-transcript}}, and LM has been ranked as the top grossing social app in the U.S. on Google Play since August 2016 and one of the top five social apps on Apple App Store. 

Both these apps share many similarities regarding community policies, e.g., they explicitly forbid broadcasters from engaging in, or broadcasting any sex-related content that promotes sexual activity, exploitation and/or assault. Moreover, both apps  prohibit violence and/or self-harm, bullying, harassment, hate speech, on-screen substance use, posting of private contact information, prank calls to emergency authorities or hotlines and solicitation or encouragement of rule-breaking. There is a variation on the user's age, as in LM users have to be at least 18 while in LL the users have to be at least 13 years old.

To counter possible violations of the aforementioned policies, both services have implemented reporting mechanisms, so that users can easily report a channel once they identify an underage user or detect suspicious behavior, or violations of the service policies. On top of that, LM employs a team of human moderators around the world, working 24/7 to respond to users' reports. Violators are subject to immediate suspension or ban from the app. Those safeguards are in place to protect young people, since live streaming apps and sites can expose them to graphic and distressing content and can leave them vulnerable to bullying and online harassment~\cite{bbc}.

However, these mechanisms do not seem to be working as intended. Many users report, for example in the app reviews, 
that they are constantly witnessing many violations of the aforementioned policies. 
It is therefore a challenge to design detection mechanisms of deviant behavior that scale up to the massive amounts of streamed video data produced in these services.
\noindent{\bf Main Contributions:} In this work we perform an in depth analysis of two SLSS to understand and characterize deviant behaviors involving the production and consumption of adult content in these platforms.
To the best of our knowledge, this is the first quantitative study of deviant behaviors in SLSS.
First, we collect two large datasets with user profile information and directed friendship links for LM and LL, following a sampling scheme that enables us to sufficiently cover the relevant part of social graphs. Next, we use a deep learning classifier to automatically identify producers of adult content from the available broadcast replays, and compare our findings with the moderation (banned users) of each platform. While our results are consistent with the moderation of LL, we observe many cases of undetected deviant behavior in LM. 
Moreover, we characterize adult content producers and consumers based on their profile attributes, and analyze their relationships to discover interesting patterns.

\subsection{Ethical considerations}
Clearly our methodology has the capacity to collect large bodies of data, including streams, messages and metadata exchanged between individuals around the world. There are therefore certain privacy considerations that must be taken into account. To anonymize users, we allocated a new unique random identifier for every user whose data we collected, obfuscating her platform-wide identity (user ID). We highlight that the terms of both services underlines that all data (and metadata for LM) and activity are by default public.
Despite their ``public'' nature, we follow Zimmer's approach \cite{zimmer2010but}. In this regard, the data remains anonymized during all the steps of our analysis, and we report only aggregated information. 
The collected datasets are publicly available online \footnote{\url{https://github.com/nlykousas/asonam2018}.}.


\section{Related Work}
As the adult content problem on SLSS has not been studied in the literature, we loosely categorize prior work into two main categories, reflecting the fundamental concepts present in this study. Finally, we provide a functional overview of the two platforms that we study.

\subsection{Social Live Streaming Services}
In SLSS users are able to stream their own live shows in real time as broadcasters, and to join the live shows of other users as viewers/audience. The audience is able to interact with the streamers through a chat and reward them with virtual rewards, e.g., points, gifts, badges (some of which are purchasable), or money. Also, various SLSS give broadcasters the opportunity to monetize part of the virtual gifts they receive from the audience during their brαodcasts. Users of SLSS employ their own mobile devices (e.g. smartphones, tablets) or their PCs and webcams for broadcasting. In contrast to other social media, SLSS are mostly synchronous \cite{Scheibe2016,Friedlander2017}, but they can also support asynchronous interactions between users, like direct messages and comments on broadcast video replays.

We differentiate between two kinds of SLSS: \textbf{General live streaming services} (without any thematic limitation), e.g. YouNow, Twitter’s Periscope, Cheetah Mobile's Live.me, (now-defunct) Meerkat Streams, YouTube live or IBM's Ustream, and  \textbf{Topic-specific live streaming services}, e.g. Twitch (games), or Picarto (art). 

Since SLSS are quite new, the literature in the field is rather limited. Some of these studies investigate the performance of such services, e.g., Meerkat and Periscope~\cite{Wang2016,Siekkinen2016},  Periscope \cite{favario2016mobile} and Twitch \cite{Deng2017}. Human factors and user experience were studied in~\cite{Tang2016}. Having access to a large dataset of Inke, a Chinese SLSS, \cite{Ma2017} identified several patterns in the users, e.g., fast interest shifts, user dedication to broadcasters as well as the locality bonds between users.

\cite{Stohr2015} analyzed traffic patterns and user characteristics of YouNow. \cite{zhao2017social} crawled Inke and identifed that the main reasons that users are hooked in these services are the follower-followee model, the awards incentivisation, and the multi-dimensional interaction between broadcasters and viewers. Similar results, but with real users, were also reported by~\cite{haimson2017makes} for the case of Facebook Live, Periscope, and Snapchat. 

Legal and ethical questions about SLSS were raised by~\cite{faklaris2016legal}. Recently,~\cite{zimmer2017law} performed an empirical study on law infringements in several SLSS. While the focus was not on adult content, the researchers found that around 17.9\% of their sample, consisting of more than $7,500$ streams, somehow violated a law, e.g., copyright, road traffic, insult, etc. Different information behaviors of users, focusing on the assessment of streamers' behavior with emphasis on produced content and motivations, as well as demographics, were studied in~\cite{Scheibe2016,Friedlander2017}. The copyright aspect is also studied in~\cite{edelman2015meerkat}, but in terms of broadcasting sport events.

\subsection{Adult content in Social Media}
In the computer science literature, adult content consumption has mostly been studied in the context of adult websites, several of which incorporate social networking functionality and features. 
Examples include the work by~\cite{Tyson2015} that provides an overview of behavioral aspects of users in the PornHub social network, a recent paper \cite{Magdy2017FakeIT} on the detection of fake user profiles in the same network, and various studies on the categorization of content, frequency of use, and analysis of user behavior in such platforms \cite{Schuhmacher2013,Tyson2013}.
To the best of our knowledge, the only other work that studies the production and consumption of adult content in general-purpose online social networks is a recent article by~\cite{Coletto2016a}. The authors perform a large-scale analysis of the adult content diffusion dynamics in Tumblr and in Flickr, while also examining and comparing the demographics of adult content producers and consumers across these platforms. A wider corpus of research has been produced by social and behavioral scientists, mostly based on surveys of relatively small numbers of individuals. 

\subsection{Live.me \& Loops Live functional overview} \label{functional_overview}
This study uses data collected from LM and LL platforms, introduced previously. 
Most of the features and functionality offered by those platforms are mobile-only, in that users wishing to actively participate in their communities need to own mobile devices such as smartphones and tablets running on  Android or iOS.

The dynamics of both communities are based mostly on three possible actions performed by the users: {\bf (a)} create real-time broadcasts and optionally associate hashtags representing thematic categories/user interests with them; {\bf (b)} join broadcasts created by other users and interact with them as well as with the other viewers. Those interactions include exchanging chat messages with other viewers, and rewarding the broadcasters with ``likes'' and purchasable virtual gifts; and
{\bf (c)} follow other users and receive notifications when they are broadcasting.

Contrary to other popular SLSS like Periscope \cite{Wang2016}, all the broadcasts in LM and LL are public. All active broadcasts are visible on a global public list. In both platforms, the concept of re-sharing/re-posting broadcasted content across different users is not present. Nevertheless, users are able to get shareable links to live shows that can be used for promoting broadcasters on other social media.

As already discussed, both platforms enable users to report community policy violators and underage users, who consequently get their accounts banned after their activity has been reviewed by moderators. Additionally, LM offers safety features to proactively protect its users, like the ``Admin'' feature, which enables broadcasters to allow other trusted users to be administrators for their broadcasts to block commenters on their behalf in real time.

Both platforms are equipped with more advanced features. Some significant examples are the ability to view currently popular/trending or ``featured'' broadcasts, either globally (both services), or by geographical region (LM), or by hashtag (both LM and LL) and the ability to find users or hashtags matching a search term. The mechanics of the broadcast featuring system are different for each platform, but in both cases factors such as the number of viewers, the amount of user interaction within the broadcast including likes, gifts and messages, and the duration of the live show are taken into account. Moreover, the popularity and experience of a user is reflected by their ``level'', which is determined by their participation in activities such as broadcasting, joining broadcasts of others, sending and receiving gifts, chatting, etc. Leveling up enables users to receive various privileges such as discounts for buying virtual currency and access to premium gifts.

Broadcasters have the incentive to get their live shows featured, since this leads to a better visibility within the app, thus attracting a higher number of viewers who in turn can potentially reward them with virtual gifts. Once a broadcaster has received a certain amount of virtual gifts, they are able cash them out for real money.  
Finally, both platforms offer a range of synchronous interaction features traditionally provided from the majority of OSNs like direct messaging between users and the ability to ``block'' users.

Follow edges in LM and LL social graphs are directed; users can follow other users who do not follow them back. In addition, following someone does not require their permission. In the context of this study we focus specifically on the user-specific attributes and following-follower social graphs of those platforms.

\section{Data collection methodology}
In this section, we first describe our methodology for collecting and labeling the data.
The objective of our data collection methodology is twofold: to identify adult content producers by analyzing the available broadcast replays and to sufficiently sample the portion of the social graphs where adult content production and consumption phenomena are predominant. To accomplish this, we develop a novel data collection and labeling approach which we detail in the following subsections.

\subsection{Sampling the social graphs}
\label{seed-selection}
Both LM and LL applications communicate with their servers using an API with SSL-protected access. 
To the best of our knowledge, no open-source clients for these services exist at the time of writing, hence, we follow a similar method as in~\cite{Siekkinen2016}. For each platform, we analyze the network traffic between the app and the service.
More precisely, we set up a so-called SSL-capable man-in-the-middle proxy between a mobile device with the specific apps installed and the LM and LL services that acts as a transparent proxy.
The proxy intercepts the HTTPS requests sent by the mobile device and pretends to be the server to the client and the client to the server,
enabling us to examine and log the exchange of requests and responses between the client apps and the servers.

\begin{figure}[t]
	\centering
	\includegraphics[width=.75\columnwidth]{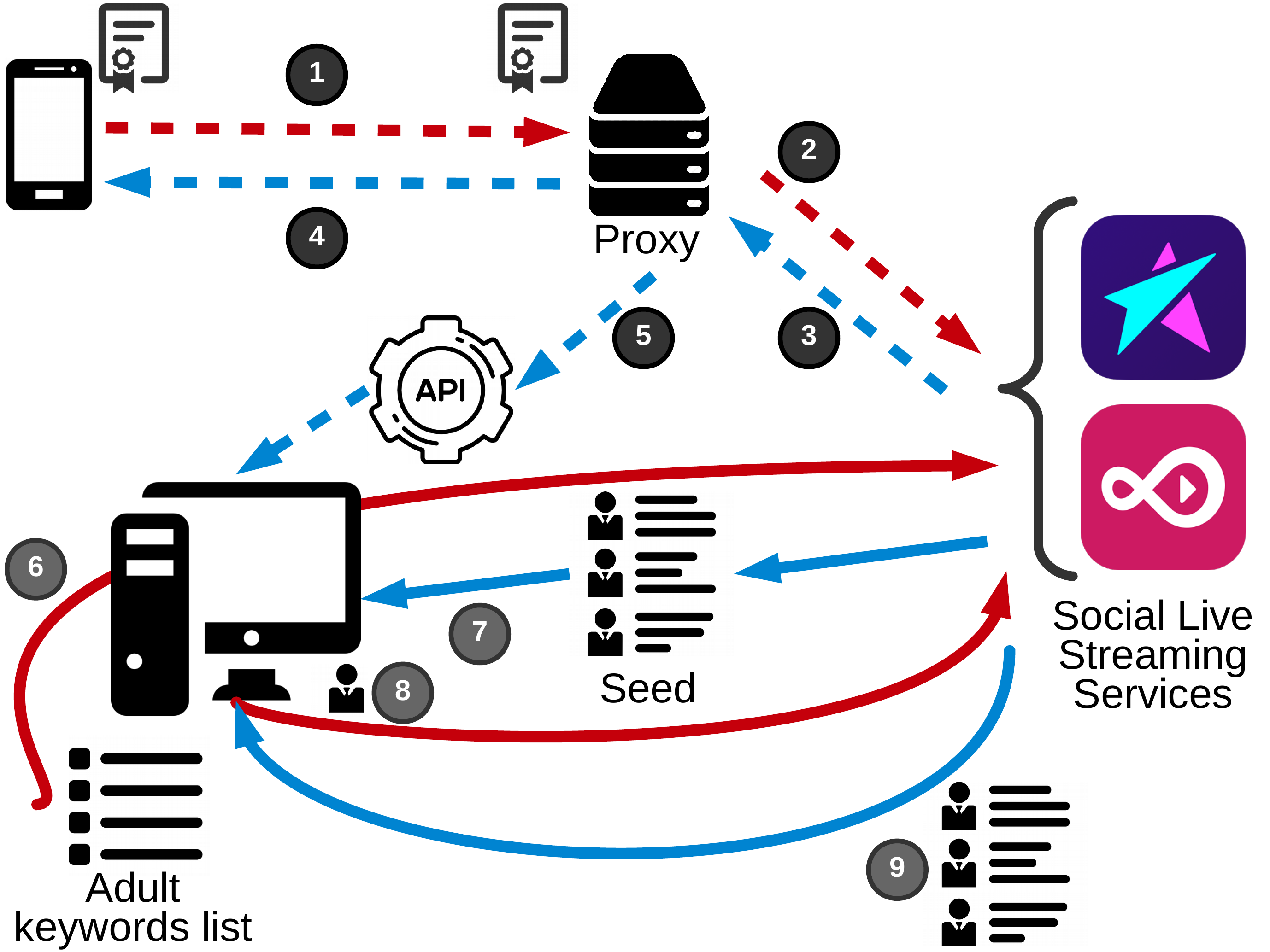}
    \caption{Data collection methodology. A proxy intercepts the messages from the smartphone to the SLSS. To decrypt the traffic and derive the API, a root certificate is installed in the smartphone (steps 1-6). Then, we use an adult keyword list to get an initial set of seed users, which are then queried to collect even more users. Based on their properties (e.g. banned, gifts) we build our dataset (steps 7-9).}
    \label{fig:intercept}
\end{figure}

We select a set of APIs that allow us to crawl the social graph edges, extract user profile and broadcast information, and use the search capabilities offered by the services. Content-wise, both services use the HTTP Live Streaming (HLS) protocol~\cite{pantos2017http} for hosting and delivering broadcast video replays, similar to other well known live streaming services such as YouNow, Periscope and Twitch.
Figure~\ref{fig:intercept} illustrates this architecture (steps 1-5).

We first identify a set of \emph{seed nodes} likely to be involved with the production of adult content, in order to bootstrap a subsequent crawling procedure for sampling the social graphs.
To accomplish this, we take advantage of the aforementioned search APIs. 
A seed node is defined as a user that satisfies \emph{the three following conditions}:~\emph{(i)}~having a username that contains a pornographic term,~\emph{(ii)}~having broadcasted activity, and~\emph{(iii)} being banned by the system\footnote{In both services, although the accounts of banned users are deactivated, their past activity in the platform is still retained, thus enabling us to perform the described analysis.}. 
For the first condition, we use the list of adult keywords provided by~\cite{Coletto2016a} in the context of their proposed deviant graph extraction procedure. 
This list contains $5,283$ search keywords from professional adult websites.

Using these three criteria, we were able to identify $390$ and $47$ seed nodes for LM and LL, respectively. 
Figure~\ref{fig:intercept} (steps 6-7) illustrates the seed identification step.
Note that this step does not consider the network structure. 
The next step (denoted as 8-9 in Figure~\ref{fig:intercept}) consists in traversing 
and collecting profile information as well as broadcast video replays from each user, following the friendship links.
We follow a Breadth-First (BF) traversal limited to two hops away (undirected distance) from the seed nodes.
Thus, our network consists of the union of the 2-hop ego-networks of all seed nodes. This union resulted in one single connected component in both platforms.
For computational reasons, we discard those nodes in the boundary that appear as neighbors of a node with degree higher than $10K$. These nodes correspond to only $718$ and $267$ profiles for LM and LL, respectively, a very small proportion of the complete 2-hop ego-networks.

We emphasize that our interest is not in capturing the entire network of users, but a tractable subset of tightly connected groups of users in which adult content is predominant.
BF search covers satisfactorily small regions of a graph \cite{Kurant2011} and has been used in many analyses. 

During the data collection period, which lasted from Jan. to Nov. of 2017, we repeated this crawling procedure once per week on average.
Based on the number of installations reported by LM on Google Play ($20M-50M$ installations), we managed to crawl roughly $5.8\%-14.5\%$ and $5.46\%-27.3\%$ of the entire LM and LL networks, respectively.

Table~\ref{tb:crawled} summarizes the obtained networks for both platforms.
As expected, the LM network is much larger than the LL network, containing approximately $10$ times more users and $30$ times more edges.
The LL network is, however, one order of magnitude more dense than the LM one.

\begin{table}
	\centering
    \caption{Network statistics of the crawled graphs: number of nodes $|N|$, number of edges $|E|$, number of banned nodes $|B|$, average degree $\langle k \rangle$, density $D$, and reciprocity $\rho$.}
     \label{tb:crawled}
      \resizebox{\columnwidth}{!}{
      \begin{tabular}{lrrrrcr}
\toprule
 & $|N|$ & $|E|$ & $|B|$ & $\langle k \rangle$ & $D$ & $\rho$ \\ \midrule
LM & 2,942,407 & 37,440,992 & 142,345 & 25.4 & 4.32 $\times 10^{-6}$ & 0.14 \\
LL & 273,177 & 1,193,780 & 114 & 8.73 & 1.59 $\times 10^{-5}$ & 0.08 \\ \bottomrule
\end{tabular}
}    
\end{table}

The approach we followed has three main limitations. Firstly, the set of replays that we captured includes only the available replays of past broadcasts at crawling time.
Replays that were deleted in-between our crawls as well as all live broadcasts streamed during our crawls were not included. 
This is not a fundamental limitation, and can be fixed by using more sophisticated approaches~\cite{Wang2016}.
Moreover, while we can determine whether an account is banned (suspended) or active, none of the platforms provides metadata to determine the reason behind the account suspension.
This means that our dataset includes false positives that were banned because of other unrelated policy violations.
This limitation is addressed in the next subsection, in which we consider the replay' content to determine whether a user is deviant or not. 
Finally, there is a small probability of false negatives, a portion of deviant users that are not retrieved by our method. This can happen because moderators can only identify a limited number of users engaging in inappropriate behavior~\cite{Cheng2015} and those may lie isolated (more than two hops away) from the seed nodes.

\subsection{Labeling the users} \label{labeling}
Having described our procedure to identify an adult content related network, we now describe how we label the users within this network.
We differentiate between three types of users: adult content producers, or simply \emph{producers} (based on their broadcast activity), \emph{consumers} (based on their relation with producers), and \emph{normal} users that are not included in any of the two other categories.


One could argue that these consumers are lurkers. The lurking phenomenon in social networks has been studied in great depth~\cite{tagarelli2014lurking,perna2018identifying}. In general, lurkers are passive users who do not contribute to the community. While consumers in our scenario could also be lurkers, we argue that, despite the obvious resemblance, they are not. Indeed, their behavior seems passive as they do not create content. Nevertheless, they have actual interactions by, e.g., providing praise and currencies to producers, or by publicly chatting with the producers, that promote specific behaviors and content.

Given the network, our criterion to establish whether a user is a producer is exclusively based on the images of the user's broadcast activity. This choice disregards indirect sources of information and does not require manual inspection, allowing us to scale up the method efficiently. Alternative approaches are based on manual inspection of metadata only~\cite{Coletto2016a}, which may not been sufficient for our purposes, or using crowdsourcing approaches for categorizing broadcasts~\cite{Friedlander2017}, which would require a pool of crowd workers to be potentially exposed to offensive material.

To this end, we use OpenNSFW\footnote{\url{https://yahooeng.tumblr.com/post/151148689421/open-sourcing-a-deep-learning-solution-for}.}, a deep neural network model pre-trained to detect pornographic images. Convolutional Neural Networks are the state of the art in image classification problems~\cite{krizhevsky2012imagenet,he2016deep}. OpenNSFW takes an image as input and provides a value representing confidence in an image's resemblance to pornography. We feed the network with frames sampled from the broadcasts at $1/3$ Hz, and keep the \emph{highest} confidence score for every broadcast replay.
This value represents the maximum probability a replay contains pornographic content.
Then, at the end of our data collection period, we can associate every user with the \emph{highest} value provided by OpenNSFW over all of their replays in the dataset.
The aforementioned value can be considered as a user's \textit{adult content production score}. For the users we were unable to collect any broadcast data, we set this value to zero.

\begin{figure}
\centering
	\includegraphics[width=.5\columnwidth]{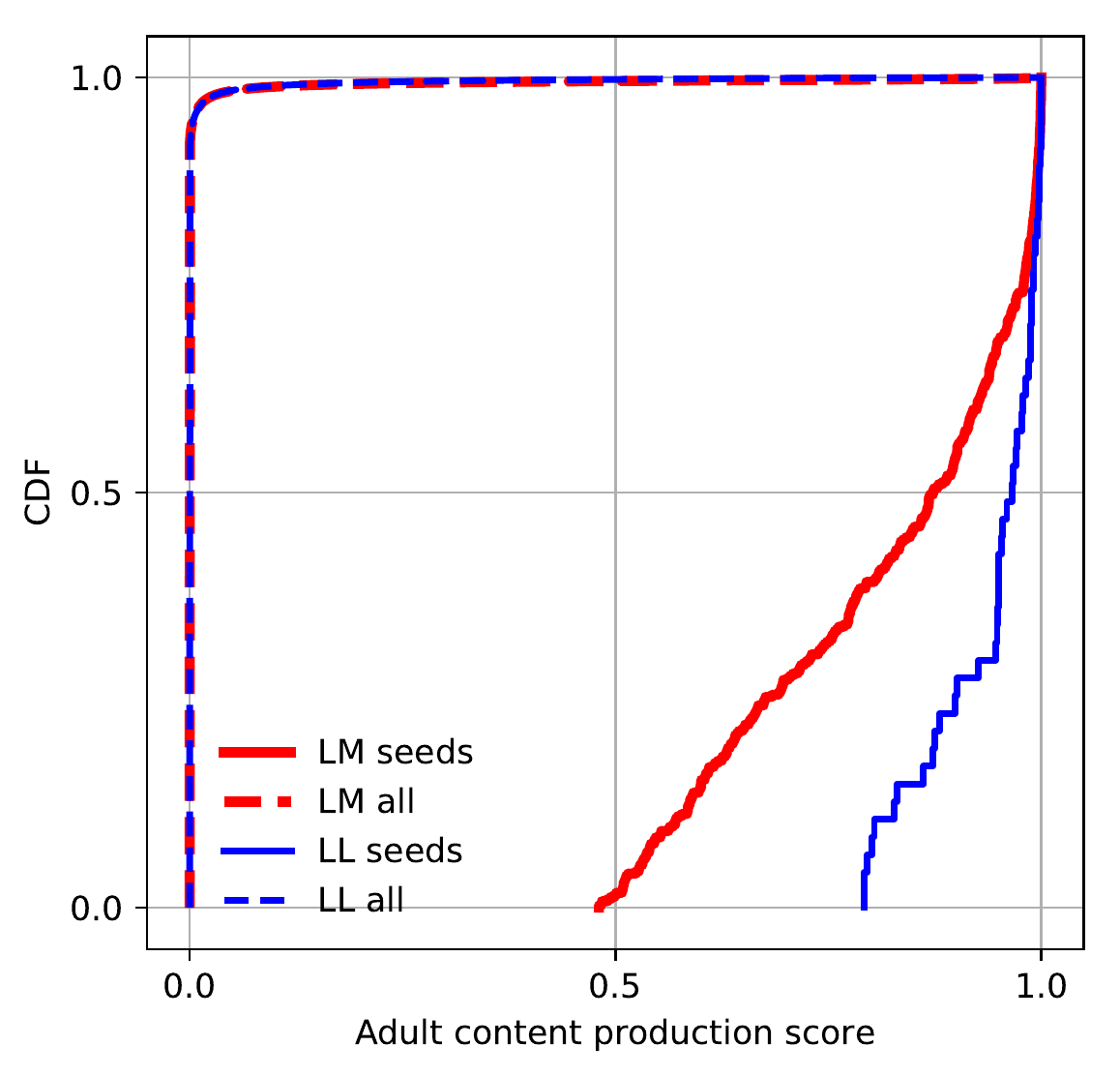}
    \caption{Cumulative distribution function of the adult content score values.}
    \label{fig:adult_scores_cdf}
\end{figure}
Figure \ref{fig:adult_scores_cdf} shows the cumulative distribution function (CDF) of the adult content production score for both LM and LL networks differentiating between seed users and all the users in our sample.
We observe that a very small proportion of all users (only around $0.4\%$) scored above $0.5$, indicating that the vast majority of users do not broadcast adult related material.
On the other hand, the seed users have been assigned high scores (all starting at $0.5$ for LM and around $0.8$ for LL). This confirms that our choice of seed nodes and the outcome of the classifier agree to a large extent.

We establish whether a user is a \textit{producer} using a thresholding approach. In particular, we consider the probability distribution of the scores for both banned and non-banned users and choose as a threshold the Bayesian decision rule that separates the two classes. 
This results in a threshold of $0.82$ and $0.93$ for LM and LL, respectively.
Although the described approach does not require any human supervision, in order to further evaluate the ``goodness" of the threshold, we manually inspected the frames 
for 100 LM and 50 LL random producers. All of them contained either nudity or semi-nudity, suggesting the validity of our thresholding method.

We establish whether a user is a \textit{consumer} based on the set of producers and the network structure.
In particular, we label a user as a consumer if the user follows \emph{at least two} adult content producers.
While following a single user (producer or not) can be expected by random chance, following two users of the producer class (given they only make up for a minor fraction of the total users), is much less likely to be by chance. 
Our definition of consumer is stricter than the one of~\cite{Coletto2016a}, which defines as a \emph{passive consumer} a user that follows \emph{at least one} single producer.
In our analysis, for those users that fall in both categories, i.e., producers that also followed at least 
two other producers, the producer label is considered more relevant. In practice, only $9$ users of LM and $2$ users of LL should have been labeled as both producers and consumers.
Finally, users who do not fall into the above classes are labeled as \emph{normal} users.

Table \ref{tab:class-dist} summarizes the resulting labeling according to our proposed procedure.
As expected, we observe that only a small proportion of the crawled networks are not labeled as normal users.
We also show in parenthesis how the seed nodes are distributed in the three categories.
Recall that seed users are banned users with broadcast activity and with a adult-related username. 
Although most of them are labeled as producers, there is a significant proportion labeled as normal users. This can be explained by the fact that those users may exhibit other (non adult-related) deviant behaviors and thus not relevant for our analysis, or because their score did not reach our threshold, as reported from the OpenNSFW classifier.
Remarkably, none of them are labeled as consumers, which already suggests that producers are not well connected between them.

\begin{table}
\centering
\caption{Distribution of users according to their class.}
\label{tab:class-dist}
	\begin{tabular}{lrr}
    	\toprule
    	{\bf Class} & {\bf Live.me} & {\bf Loops Live}\\
        \midrule
        Producers & 7,135 (228 seeds) & 92 (33 seeds)\\
        Consumers & 30,872 (0 seeds) & 1,243 (0 seeds)\\
        Normal & 2,904,400 (162 seeds)& 271,842 (14 seeds)\\
        \bottomrule
    \end{tabular}
\end{table}

\subsection{Effectiveness of SLSS moderation systems} \label{moderation}
Having identified the aforementioned user classes, we proceed to examine how our labeling approach compares to the moderation of each platform.
Table~\ref{tab:moderation-confusion} shows how banned users are distributed in each class.
\begin{table}[!b]
	\centering
    \caption{Proportion (total in parenthesis) of banned accounts in each class.}
    \label{tab:moderation-confusion}
	\begin{tabular}{lrr}
    	\toprule
    	{\bf Class} & {\bf Live.me} & {\bf Loops Live}\\
        \midrule
        Producers & 43.5\% (3,109) & 96.7\% (89) \\
        Consumers & 9.6\% (2,970) & 0.08\% (1) \\
        Normal & 4.6\% (136,266) &  0.008\% (24) \\
        \bottomrule
    \end{tabular}
    
\end{table}
In the case of LM, we observe that only 43.5\% of the labeled producers have been banned.
Since it is unlikely that the frames extracted from the broadcasts contained adversarial perturbations~\cite{Szegedy2013} against the OpenNSFW model,
we can safely assume that moderation is highly ineffective in detecting such cases.

On the contrary, moderation of LL is consistent with our labeling outcome, with 96.7\% of users placed in the producers class being banned. This consistency provides further confirmation of our decision to use a pre-trained deep learning classifier for detecting adult content.  Finally, the high number of banned users placed in the \textit{normal} class suggest the existence of a significant proportion of policy violators outside the context of our study.

\section{Profiling deviant users}
In this section, we present our efforts to characterize adult content producers, consumers and their relationships in the sampled networks.
We first consider a set of features directly accessible from each user and analyze their relevance for distinguishing between classes: normal users, producers, and consumers. 
We then look at the network structure to gain understanding about the relations between consumers and producers.

\subsection{Features} \label{sc:features}
Based on the available profile information we collected from the two platforms, we define a set of features that can be grouped as follows:
\begin{compactitem}
\item  \textbf{Network features}:
	, Number of followers. 
	, number of followings. 
	, number of bidirectional friends. 
\item  \textbf{User-based features}
	\begin{compactitem}
	\item Pornographic username (binary): \textit{whether  the username contains a pornographic term.}
	\item Suspended/Banned (binary): \textit{whether the account has been suspended by platform moderators.}
	\item Replay count: \textit{Number of past broadcasts available for replaying.}
	\item Level: \textit{An integer value reflecting the participation level of a user in various SLSS-specific activities.}
	\item Praise (\textbf{only LM}): \textit{Total number of likes received in all user's broadcasts.}
	\item Income (\textbf{only LM}): \textit{Total virtual currency value of gifts received in all of user's broadcasts.}
	\end{compactitem}
\end{compactitem}
We assessed the relative power of these features in discriminating the three user classes by using the Mean Decrease Impurity (MDI) metric, where a higher score implies a more important feature.
Table \ref{tab:feature-importance} reports the ranking of the top five most important features in differentiating the three user classes 
for each platform.

\begin{table}
\centering
\caption{Top 5 features for differentiating the three classes.}
    \label{tab:feature-importance}
\begin{tabular}{llc|lc}
\toprule
 & \multicolumn{2}{c}{\textbf{LM}} & \multicolumn{2}{c}{\textbf{LL}} \\ 
\multicolumn{1}{c}{\textbf{Rank}} & \multicolumn{1}{c}{\textbf{Feature}} & \textbf{MDI} & \multicolumn{1}{c}{\textbf{Feature}} & \textbf{MDI} \\\midrule
1 & \#Followings & 0.31 & \#Followings & 0.37 \\
2 & \#Followers & 0.25 & \#Friends & 0.26 \\
3 & Praise & 0.15 & \#Followers & 0.19 \\
4 & \#Friends & 0.12 & Banned & 0.10 \\
5 & Income & 0.06 & Porn nickname & 0.05 \\ \bottomrule
\end{tabular}
\end{table}

\begin{figure*}[!ht]
  \centering
  \begin{subfigure}[t]{0.19\textwidth}
  	\centering
    \includegraphics[width=\textwidth]{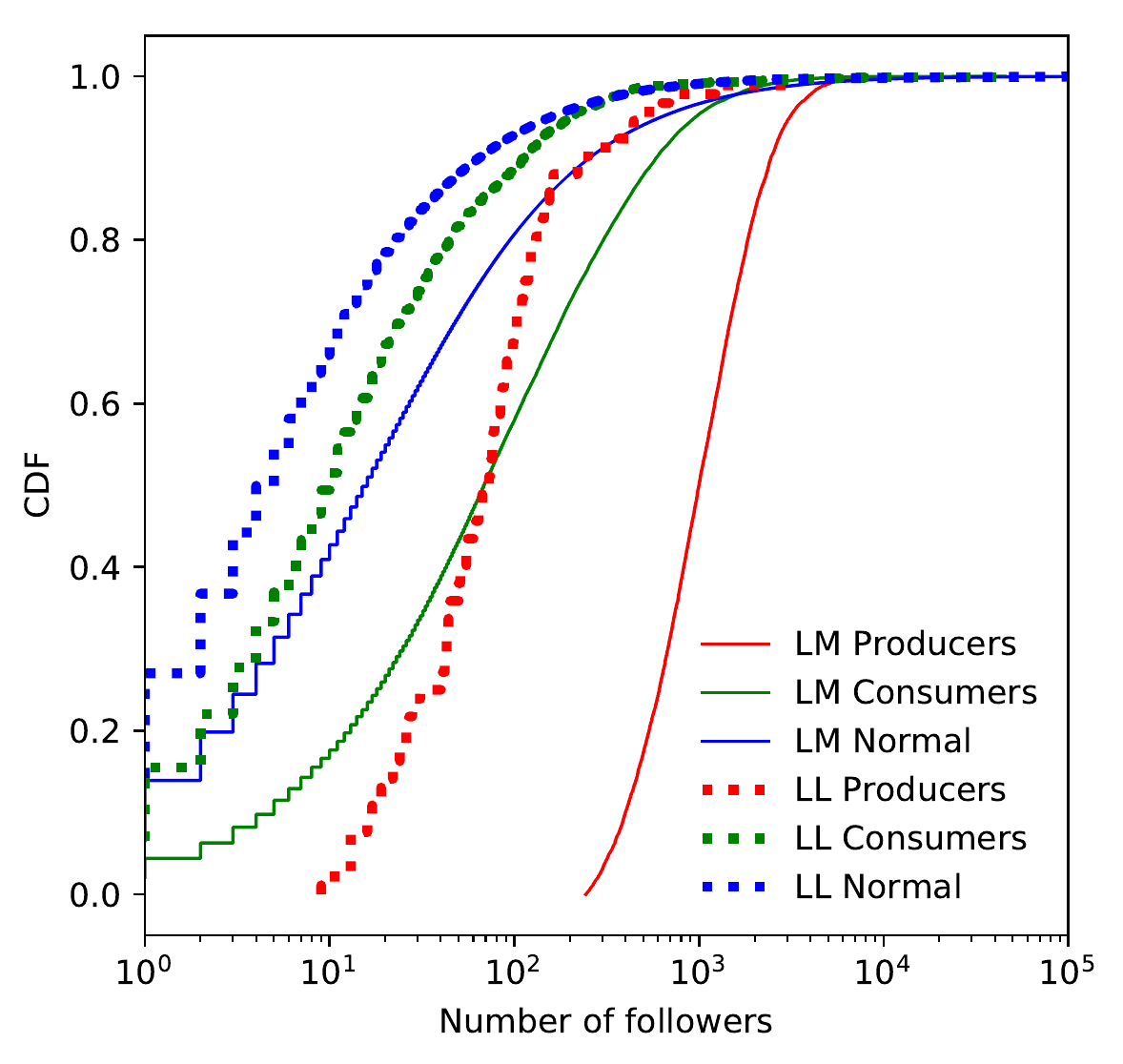}
    \caption{\# Followers}
    \label{fig:followers_cdf}
  \end{subfigure}
  \begin{subfigure}[t]{0.19\textwidth}
  	\centering
    \includegraphics[width=\textwidth]{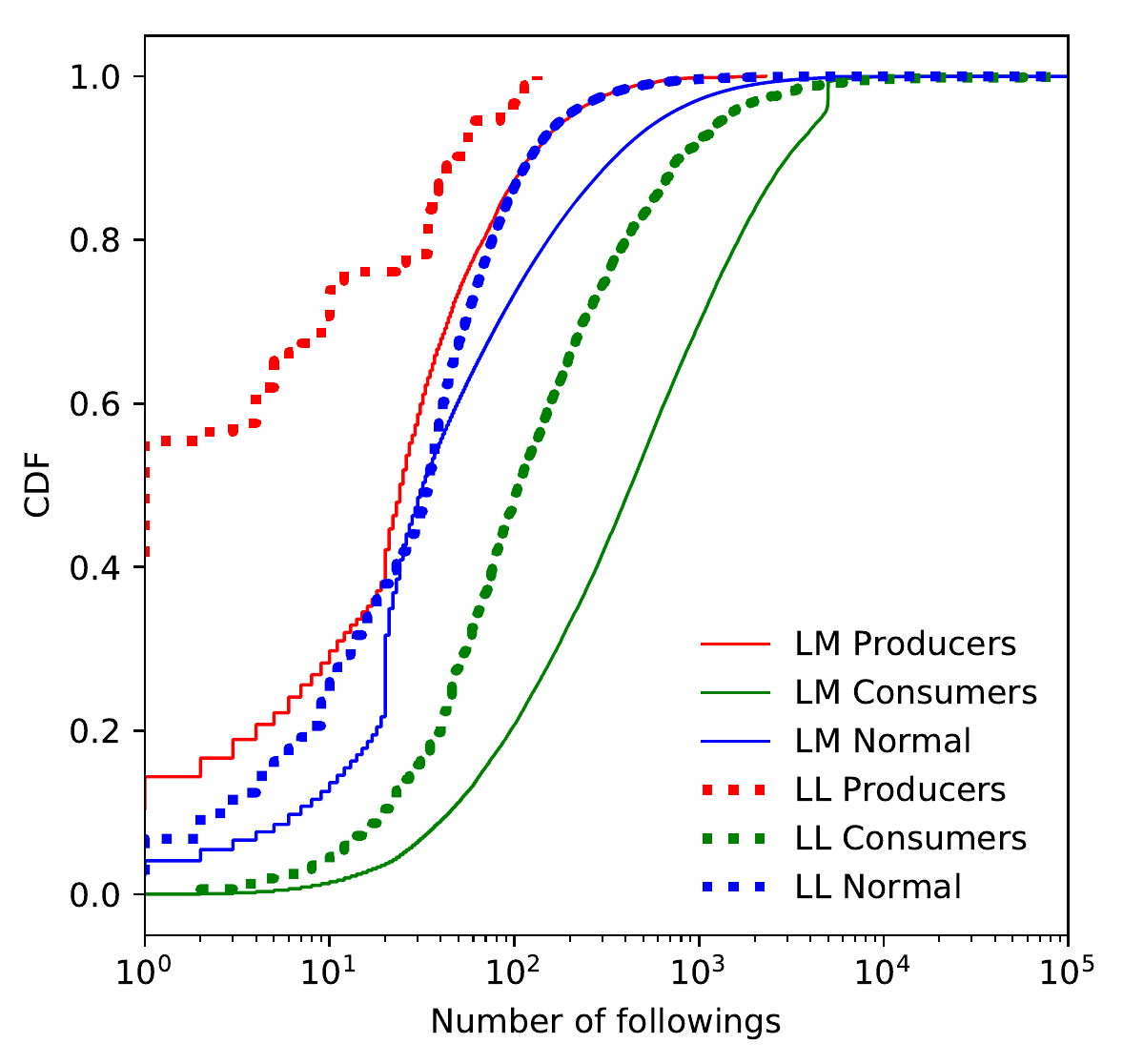}
    \caption{\# Followings}
    \label{fig:followings_cdf}
  \end{subfigure}  
  \begin{subfigure}[t]{0.19\textwidth}
  	\centering
    \includegraphics[width=\textwidth]{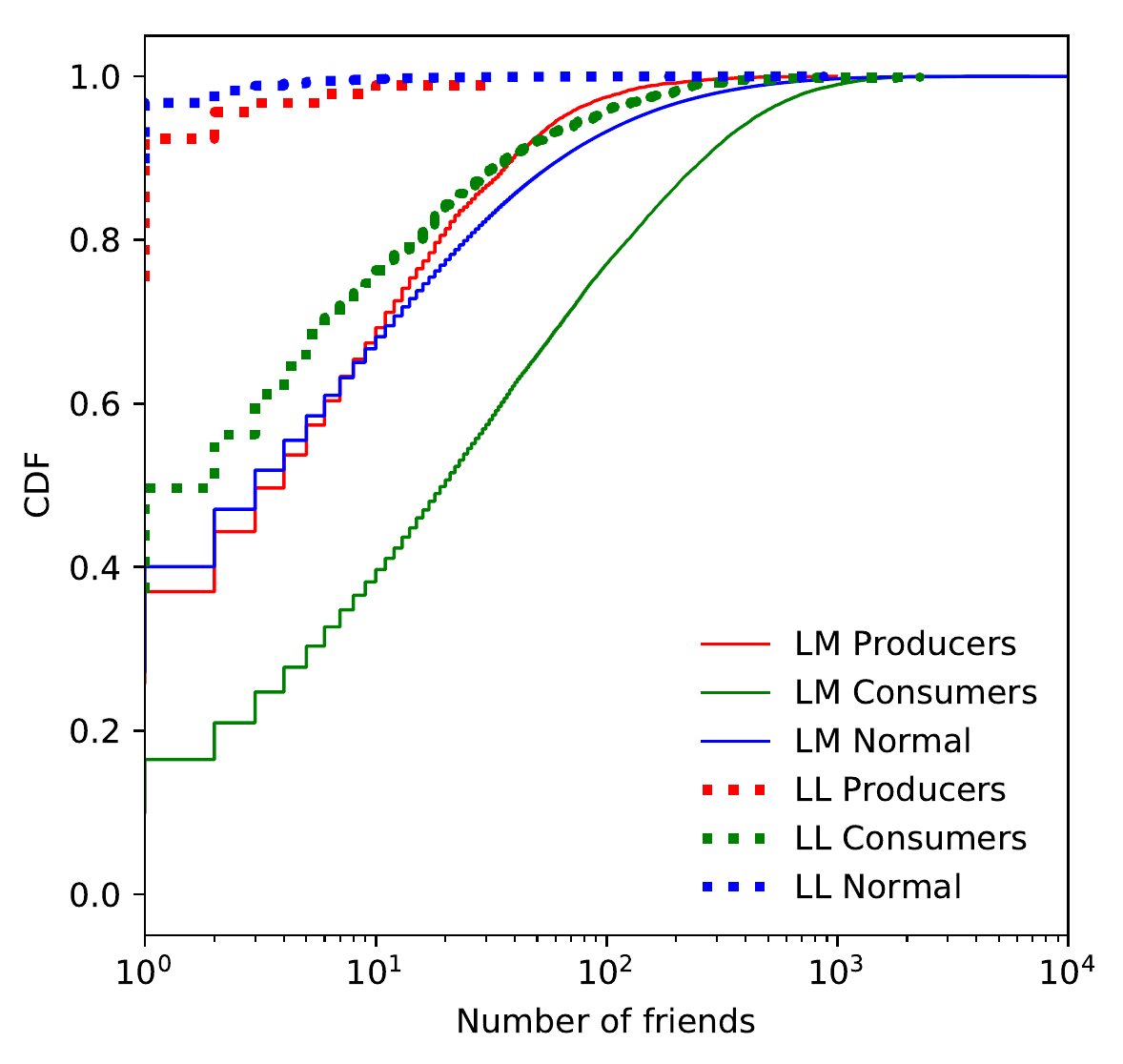}
    \caption{\# Friends}
    \label{fig:friends_cdf}
  \end{subfigure}  
  \begin{subfigure}[t]{0.19\textwidth}
  	\centering
    \includegraphics[width=\textwidth]{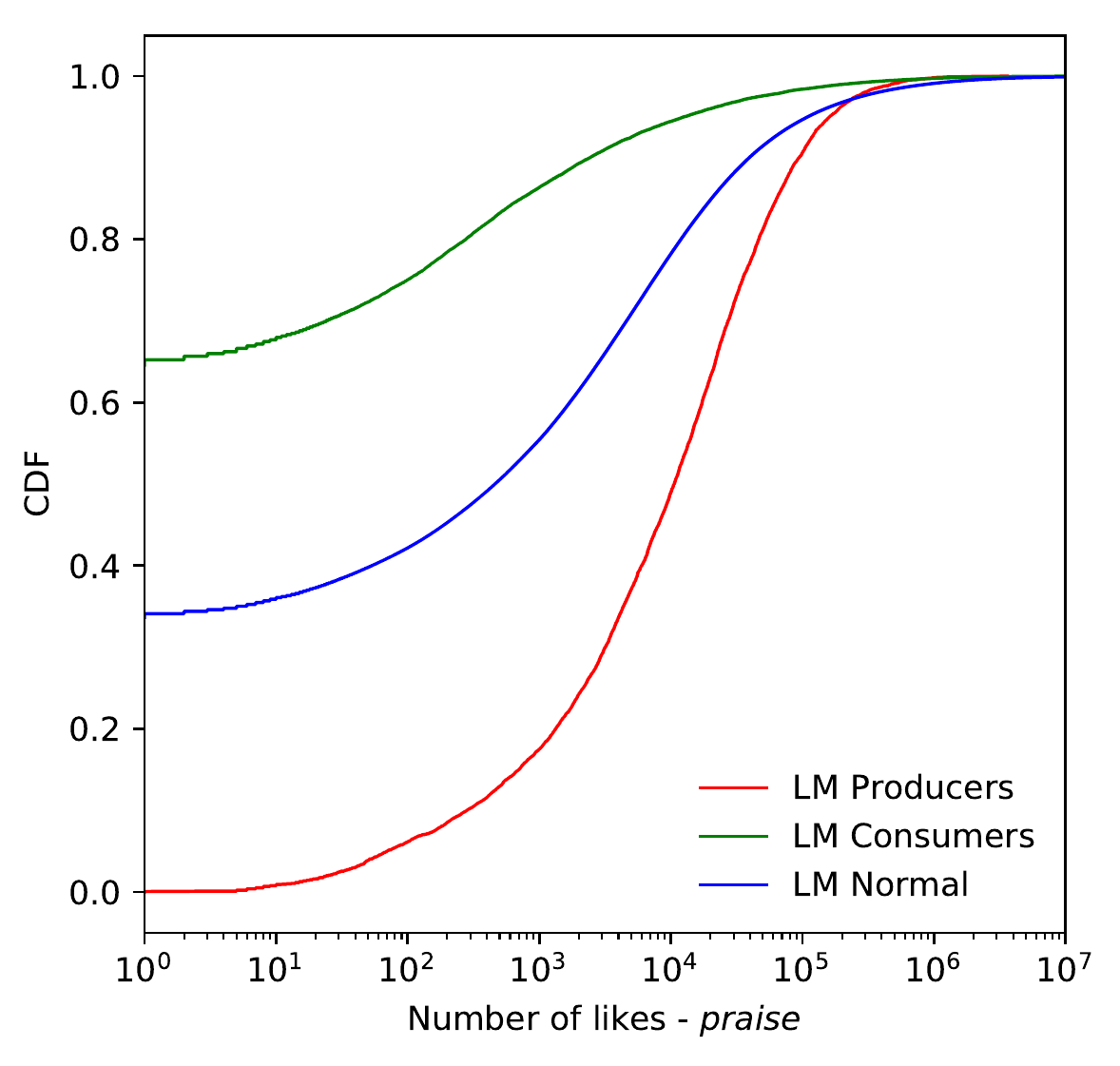}
    \caption{Praise}
    \label{fig:praise_cdf}
  \end{subfigure}  
  \begin{subfigure}[t]{0.19\textwidth}
  	\centering
    \includegraphics[width=\textwidth]{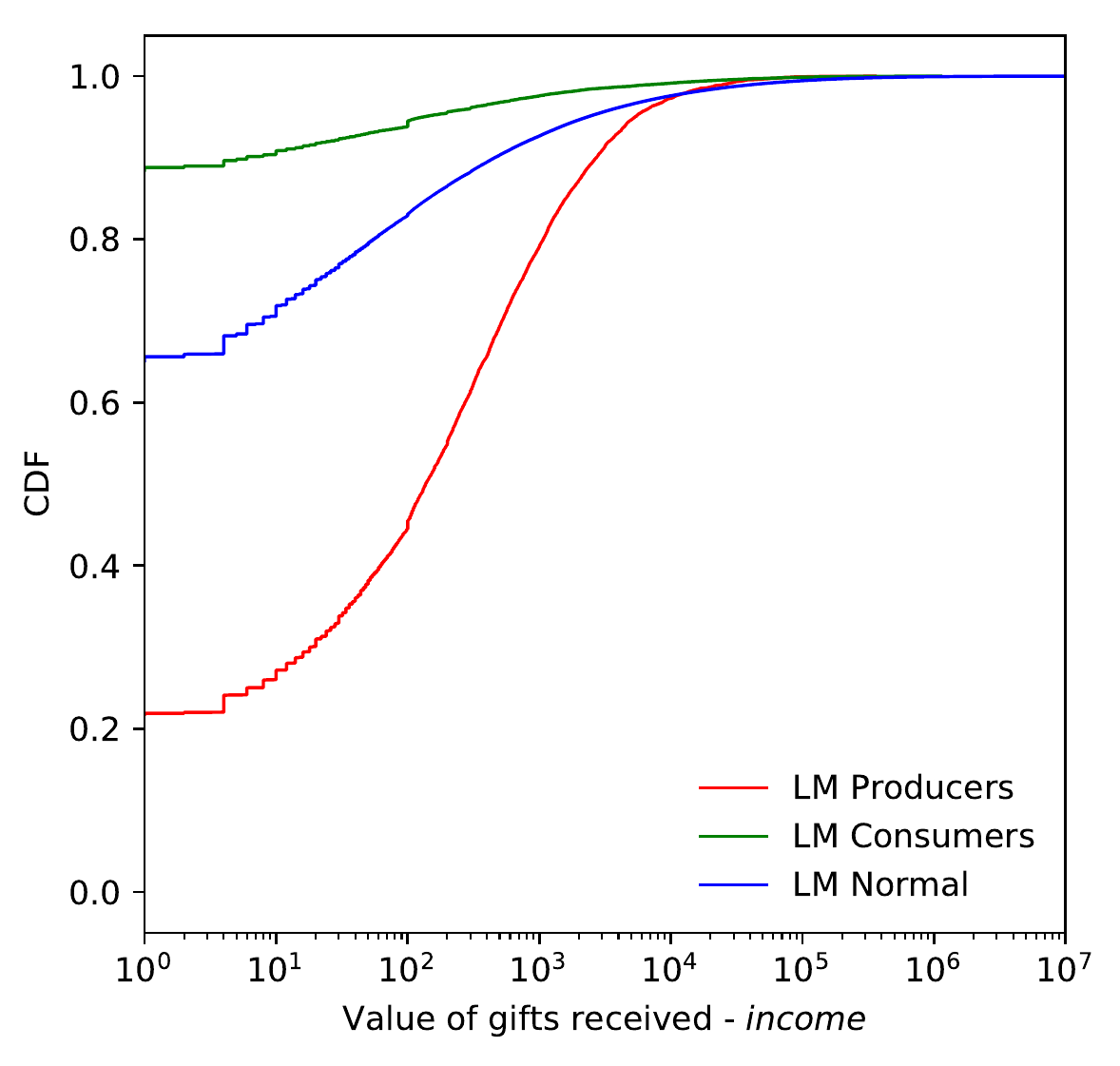}
    \caption{Income}
    \label{fig:income_cdf}
  \end{subfigure}  
  \caption{Cumulative distribution functions (CDFs) of the different user profile attributes.}
  \label{fig:feature_importance_cdfs}
\end{figure*}

The number of followings, followers and friends are among the highest ranked features for both networks, which suggests relevance of social relationships for characterizing the given classes. Also, for LM we observe that the amount of likes (praise) and virtual gifts (income) are highly important as well. To get a deeper insight on how these features are distributed across the different classes, we plot their cumulative distribution functions (CDFs) in Figure~\ref{fig:feature_importance_cdfs}. Information about praise and income was not available for LL, preventing us from performing an 1-to-1 comparison between the two datasets. Instead, we observe a high importance for the banned and pornographic username attributes. This is due to the fact that, as shown in Subsection~\ref{moderation}, the banned LL users are almost exclusively adult content producers, and also a significant part of them have pornographic usernames (see Subsection~\ref{seed-selection}).

From Figure \ref{fig:followers_cdf}, we observe that adult content producers tend to have many more followers than the other classes, and there exists a lower bound to the follower number of producers, approximately $10$ and $250$ for LL and LM, respectively. In contrast, consumers in LM are found to have the least amount of followers among the three classes. For the number of followings (Figure \ref{fig:followings_cdf}), however, the situation is reversed. Consumers dominate over the other classes by following significantly higher amounts of users, while the producers come last in this aspect with around $41\%$ (LL) and $10\%$ (LM) of them not following any other users. The distribution of the friend number reveals that consumers are much more likely to form reciprocal relationships, while it appears to be almost identical for producers and normal users, as Figure \ref{fig:friends_cdf} indicates. Furthermore, we found that adult content producers tend to receive the highest amount of praise and income among the three classes. 
We note that, although the higher (undirected) degree of consumers and producers is explained by the criteria used in the seed selection, the edge directionality can not be fully attributed to our sampling method, which is blind with respect to it.

Additionally, consumers receive much less recognition for their broadcasting activities compared to both normal users and producers. In fact, while no producers with zero praise exist, approximately $65\%$ of consumers and the $33\%$ of normal users in our dataset fall in this ``unpopular'' category, see Figure \ref{fig:praise_cdf}. This either means that they have not received any likes during their shows, or they have never broadcasted anything. A similar trend is observed for the total value of the virtual gifts received, represented by the income attribute and illustrated in Figure~\ref{fig:income_cdf}. Only $21\%$ of producers have not received gifts, while the same holds for the $88\%$ of consumers and the $65\%$ of normal users.

\subsection{Deviant relationships}
To determine the community structure of the these networks, existing variants of the Louvain method~\cite{Blondel2008} do not find well identifiable clusters of users.
In both networks, we observe that producers and consumers are distributed nearly uniformly across the clusters.
Further, the results vary significantly between different runs. 
We thus adopt a different approach in order to better understand the network structure. 

\begin{figure}
   \begin{subfigure}[t]{0.49\columnwidth}
	\centering
      \includegraphics[width=.9\textwidth]{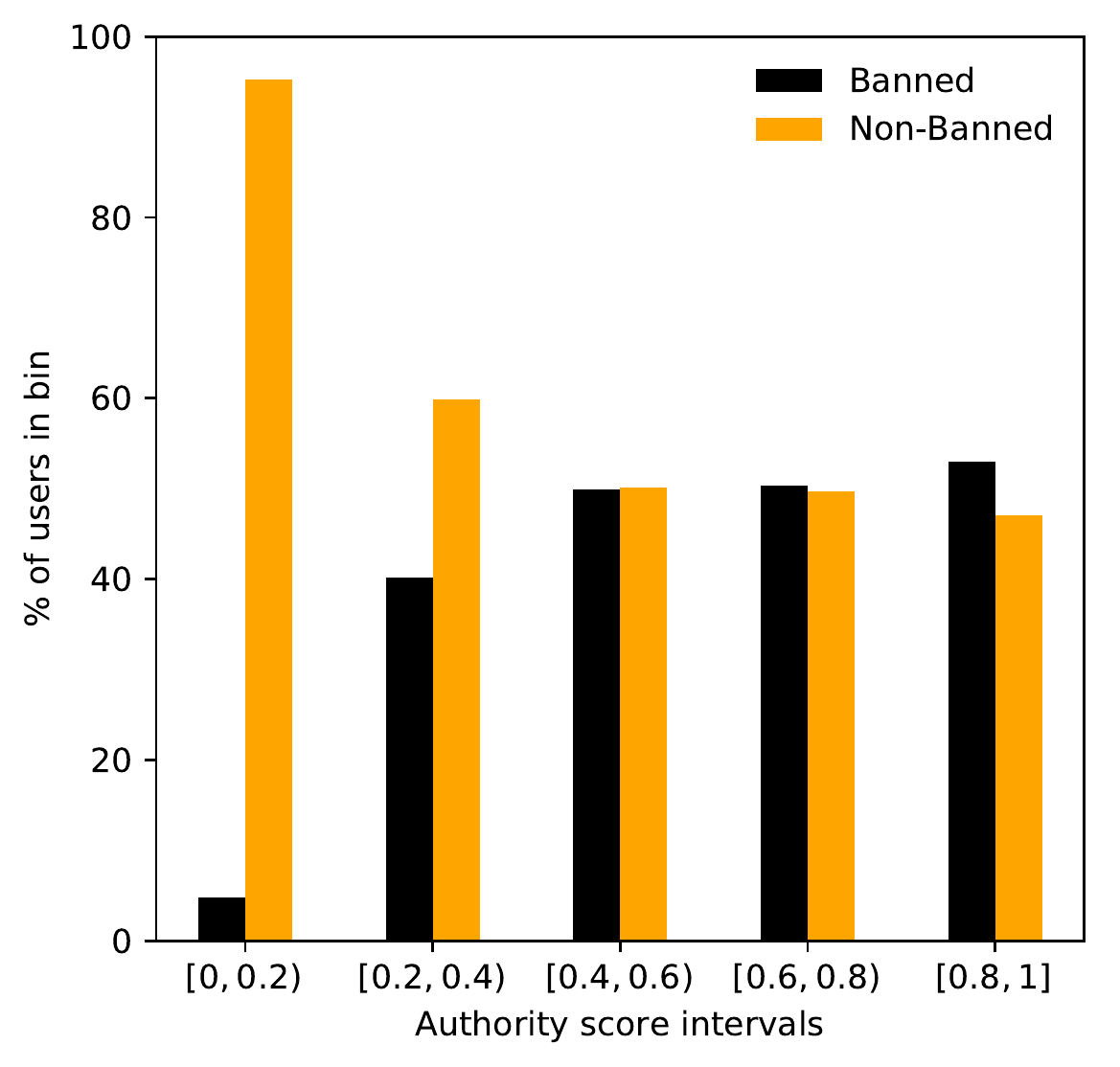}
      \caption{authority score bins (LM)}
      \label{fig:auth_bins}
   \end{subfigure}
       \begin{subfigure}[t]{0.49\columnwidth}
	\includegraphics[width=.9\columnwidth]{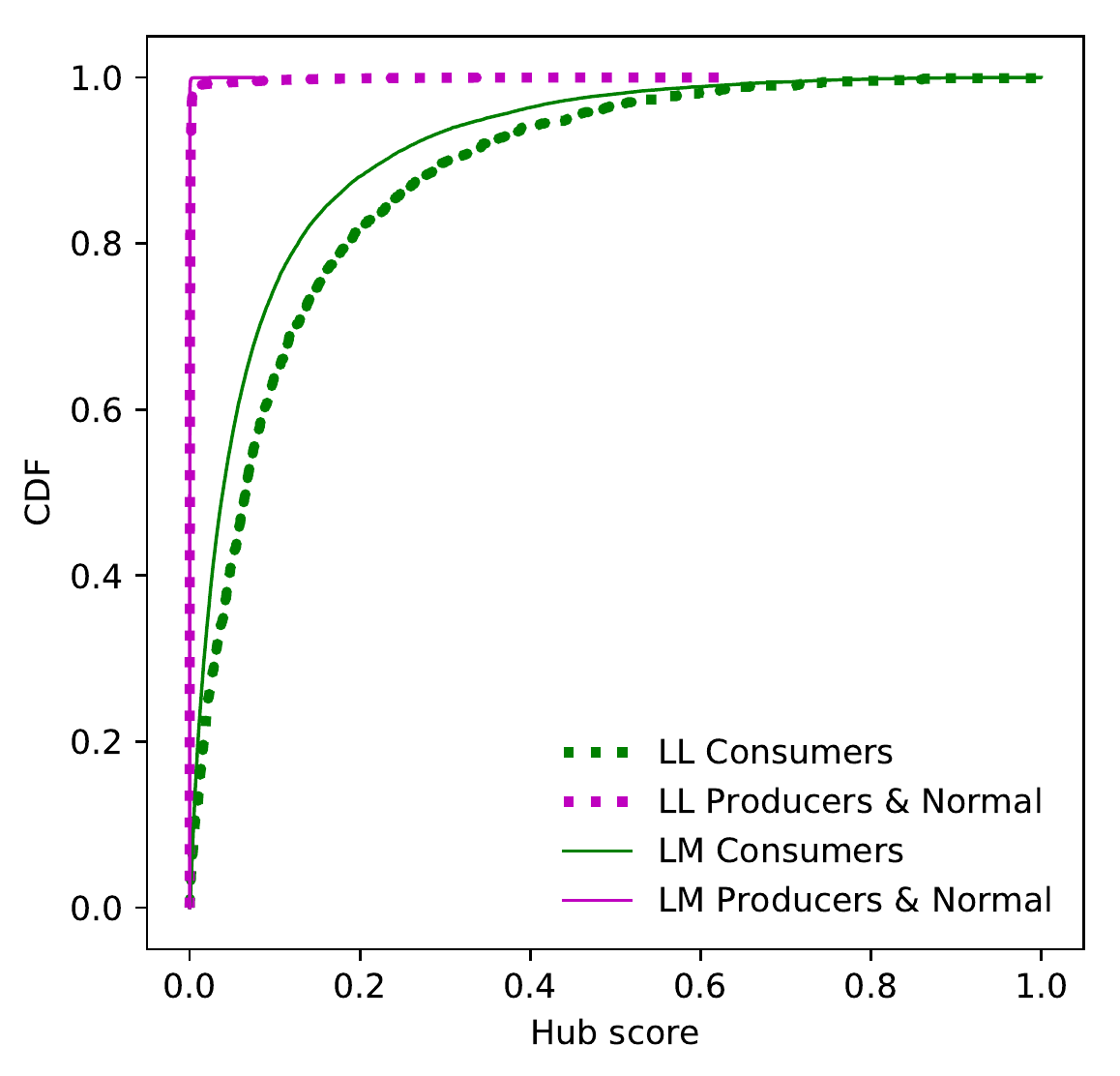}
    \caption{CDFs of hub scores}	
    \label{fig:hub_cdf}
	\end{subfigure}
  \caption{User relationship insights, provided by HITS.}
\end{figure}

In particular, we examine who in the sampled networks is significant with regards to their social relationships.
We use the ranking HITS algorithm~\cite{Kleinberg1999} to identify the hubs and authorities in the social graphs. The basic principle behind HITS algorithm is the following mutually reinforcing relationship between hubs and authorities: good hubs  point to many good authorities and vice-versa. Interestingly, it appears that adult content consumers have the highest hub scores among all users in both networks, as shown in Figure \ref{fig:hub_cdf}. Moreover, the highest authority scores in LM belong almost exclusively to producers. The latter could be correlated with the significance of the number of followers and followings to discriminate producers and consumers, from normal users, as previously shown. 
For LL, we observe that most authoritative users do not belong to the producers class, and exhibit characteristics expected of prominent users in a social community such as the number of followers in the order of hundred of thousands. The reason behind this difference between LM and LL could be attributed to the very limited extent of the adult content production behavior in the later. Therefore legitimate popular users  dominate the authority scores in the sampled graph by being followed by consumers. 
Additionally, we notice that the hub scores of the highly authoritative users are particularly low in both networks. This finding contradicts other studies on different social networks such as Twitter \cite{Leea,Java}, where researchers observed many well-connected users that have high scores as both authorities and hubs.

An interesting finding was that in LM, the ratio of banned to non-banned users increases along the increase of authority score. To better demonstrate this, we bin the users based on 
their authority score and we calculate the fractions of banned and non-banned users in each bin, as shown in Figure~\ref{fig:auth_bins}. We observe that $99.5\%$ of users in our sample fall into the first bin, having authority score less than $0.2$. We can thus conclude that banned users are more densely concentrated towards the higher end of the authority score spectrum, and the reason behind their suspension was likely the production of adult content, since they are followed by the consumers/hubs. A similar phenomenon is observed for LL, but with banned users mostly concentrated in the $0.02-0.35$ authority score range, while the $97.7\%$ of users have authority scores below $0.02$.

Based on the arguments above, we expect that consumers will follow multiple producers, a considerable portion of which will be banned. 
To quantify the relationship between the fractions of banned users and producers followed by consumers, we calculate their correlation using Spearman's rank correlation coefficient $\rho$. Indeed, there exists a nearly perfect correlation for LL with $\rho=0.96$, meaning that consumers do not follow almost any banned users outside the producers class, and a moderately strong correlation in LM ($\rho=0.63$).

Another dimension to examine is the connectivity within each class in the context of the sampled graphs. For this we measure the edge density, computed as the ratio of edges between the users belonging in each class over the total number of possible edges between them. 
 To account for the differences in sizes of the subnetworks~\cite{Leskovec2005}, we resort to a comparison of the connectivity of the sampled graphs with a null model
that randomly rewires the edges while keeping the degree of each node unchanged, as described in~\cite{Xiao2010}. 

Table \ref{tab:density} contains the link density comparison between the subgraphs induced by producers, consumers and normal users in  each sampled network and the null model.
\begin{table}
\centering
\caption{Comparison in terms of link density $D$ between the crawled networks of producers, consumers normal users and a correspoding random network.}
    \label{tab:density}
    \resizebox{\columnwidth}{!}{
    
	\begin{tabular}{llrr}
    	\toprule
    	& {\bf Class} & {\bf Crawled graph $D$} & {\bf Null model $D$}\\
        \midrule
        &Producers & $1.90 \times 10^{-5}$ & $4.55 \times 10^{-6}$\\
        {\bf LM}&Consumers & $1.20 \times 10^{-3}$ & $4.46 \times 10^{-6}$\\
        &Normal & $2.18 \times 10^{-7}$ & $4.32 \times 10^{-6}$\\
         \midrule
        & Producers & $1.91 \times 10^{-3}$ & $0$\\
        {\bf LL} &Consumers & $3.28 \times 10^{-3}$ & $1.42 \times 10^{-5}$\\
        & Normal & $1 \times 10^{-5}$ & $1.59 \times 10^{-5}$\\
        \bottomrule
    \end{tabular}
    }
\end{table}
We observe that consumers, when compared to the null model, are several orders of magnitude more densely connected to each other in the sampled networks. On the contrary, the subgraphs of producers and normal users are much more sparse compared to consumers, with the producers being only slightly more dense connected than random for LM. This finding comes in contrast with the behavior of adult content producers in Tumblr and Flickr, where they are observed to form densely interconnected communities \cite{Coletto2016a}. In LL the producers appear to have a density comparable to those of consumers, but given their limited number, this is due to the existence of producers who also exhibit consumer behavior.
 
In summary,  we can conclude that the closely knit groups of consumers act as a ``bridge'' between the otherwise isolated producer nodes. Concretely, from a network perspective, the most effective way to reach adult content in the studied networks is by traversing the social links of consumer nodes that point to both producers and other consumers, enabling the reach of even more deviant nodes belonging in those two categories.
Since adult content producers are isolated in the network, we speculate that some of the consumers are actively monitoring the list of active broadcasts (see Subsection~\ref{functional_overview}), and proceed to follow users who broadcast adult content, while also possibly sharing links to such live streams with other consumers. These ``consumer leaders'' are likely to become popular among their kin by being followed by many other consumers, thus serving as a means for diffusion of information about producers, effectively compensating for the absence of the content reposting functionality in SLSS.

\section{Discussion}

With the continuous growth of SLSS, an increase in deviant behaviors on social media is expected. In our work, we show that current moderation mechanisms may have important limitations when addressing the detection of adult content consumption and production.
Our approach overcomes scalability issues that appear when a large number of humans are needed to categorize the content, at the cost of relying on the accuracy of automatic image classification.
Image classification is the primary application domain for machine learning~\cite{Jordan255}, reaching human-level performance in many tasks.
Our results could be further improved by replacing or accommodating the OpenNSWF classifier with more effective models.

The inefficiency of moderation can be partially attributed to a {\it voyeur} phenomenon. Many adult content producers are not reported to moderators as the consumers like the content, so their accounts are not suspended, allowing them to continue broadcasting inappropriate content. 
Moreover, although consuming any kind of content, including adult, is not explicitly prohibited by the community guidelines of these platforms, suspending the accounts of the users who intentionally seek adult content would be meaningful, due to the law of  \textit{supply and demand}.
It is therefore necessary to incorporate effective, real-time detection mechanisms of deviant behaviors in the existing moderation systems, in order to maintain the SLSS communities safe, especially for the younger audience.

In future work, we will investigate quantitatively the identification between consumers and lurkers. Moreover, we plan to develop graph-based features for the detection and classification of adult content producers and consumers in SLSS by exploiting the characteristics of deviant behavior presented in this paper, as well as study other available data from broadcast-related user interactions in SLSS (chat messages, likes, gift exchange), to further analyze the nature of deviant behaviors in such platforms.

\section*{Acknowledgments}\label{sec:Acknowledgments}
We thank Andreas Kaltenbrunner for his helpful comments and suggestions.
This work was supported by the European Commission under the Horizon 2020 Programme (H2020), as part of the 
\href{http://practicies.org/}{Practicies} 
project (Grant Agreement no. 740072), and by the Spanish Ministry of Economy and Competitiveness under the Mar\'ia de Maeztu Units of Excellence Programme (MDM-2015-0502).

\balance
\bibliographystyle{IEEEtran}
\bibliography{bibliography,reference}

\end{document}